\documentclass[prd,nofootinbib,tightenlines,notitlepage,preprintnumbers]{revtex4-1}   
\usepackage{latexsym,amsmath,amssymb,amsbsy,amsfonts,graphicx}  
\usepackage{epstopdf}
\usepackage[cp1251]{inputenc}  
\usepackage[left=3cm, right=1.5cm, top=1cm, bottom=2cm]{geometry}  
\usepackage{color}  
\usepackage{verbatim}  
\newcommand{\e}{\mathop{\rm e}\nolimits}  
\newcommand{\vecr}{\vec{r}}

\newcommand{\vf}{v_{\rm F}}

\begin{document}  
\hfill HU-EP-13/31  

\title{\Large{Pseudopotential model for   
Dirac electrons in graphene   
with line defects  
}}  
\author{\large D. Ebert$^{1a}$, V.Ch.~Zhukovsky$^{2b}$, and E.A.~Stepanov$^{2c}$}  
\affiliation{$^{1}$ Institute of Physics, Humboldt-University  
Berlin, 12489 Berlin, Germany}  
\affiliation{$^{2}$Faculty of Physics, Moscow State University, \\119991, Moscow, Russia\\  
$^{a}$debert@physik.hu-berlin.de, $^{b}$zhukovsk@phys.msu.ru, $^{c}$stepanov@physics.msu.ru}

\begin{abstract}  
We consider electron transport in a planar   
fermion  
model containing various types of   
line defects modelled by   
$\delta$--function pseudopotentials with different matrix coefficients. The  
transmission probability for electron transport through the defect line  is obtained for various types of pseudopotentials. For  
the schematic model considered  
that may describe a graphene structure with different types of linear
defects, the valley   
polarization is obtained.  
\end{abstract}  
\maketitle  
 
 
\section{Introduction}  
 
For the last years special interest in 2+1 dimensional models appears  
in condensed matter physics. An  
important prototype of such  models is graphene \cite{Novos},  
\cite{1}, \cite{2},   
a planar monoatomic layer of carbon,   
which may be regarded as a  
superposition of two triangular sublattices, $A$ and $B$,  
forming a hexagonal lattice.  
As it has been recently discovered, graphene posesses various unusual  
properties. For instance, in  \cite{Geim10}, \cite{Geim0}  such  
properties as anomalous Hall effect, conductivity and other  
interesting features of material were investigated.  
The behavior of electrons in problems related to graphene  
can be effectively described by the  Dirac equation for massless  
fermions obtained from a continuum version of the tight-binding model  
\cite{wallace}, \cite{Semenoff},  
\cite{Gus}, \cite{Neto2}.  
A chiral gauge  
theory for graphene was formulated in \cite{jackiw}. Further studies  
of the theory of two-dimensional tight-binding quantum systems,   
as described  
in the continuum approximation  
by the Dirac equation in (2+1)-dimensional  
space-time   
with account  
for topological properties,   
were made in \cite{Hou},\cite{Chamon}, \cite{Obispo}.  
Note that, despite its similarity, this equation is in this context  
not a relativistic wave equation, but  
arises by  
linearizing the energy as a function of a momentum near   
Dirac points, i.e. intersections of the energy dispersion with  
the Fermi level.   
 
Recently,  models  
with different types of defects in the structure of planar systems  have  
attracted much attention. These defects can lead to  
many nontrivial properties of the transmission of propagating particles,  
which are     
related to nonuniform  
densities near the defects and barriers.  
Recent investigations in graphene provide various  examples of  
this kind of problems.  
Let us mention, in particular, the   
recently observed     
topological   
line  
defect, containing    
the periodic repetition of one octagonal plus two pentagonal carbon rings  
along a certain direction    
embedded in a perfect graphene sheet  
\cite{Lahiri}, and also interesting grain boundaries \cite{Huang} in  
graphene.     
Clearly,   
more new important applications of low-dimensional structures can be realized, when    
transport    
problems  
in them  are well understood.   
In particular,  
as   
line   
defects have a simple geometry, this makes     
them  
easier for a  
theoretical study and   
suitable for the use for controlled transport in  
graphene.     
In this context, let us mention the   
recent theoretical studies of electronic transport through a  
line defect in graphene     
considered in \cite{China2}, which  
were based upon the Green function approach.      
 
The aim of this paper   
is to study   
line defects as  barriers for the    
electron  
propagation   
by using the  effective  
Dirac  equation for   
massless   
electrons   
in (monolayer) graphene   
\cite{wallace,Semenoff,Gus,Geim10,   
Geim0},   
in the  
framework of a  schematic {\it pseudopotential model}. We   
shall  
consider  all     
possible    
types of {\it barrier-type}   
perturbations, chosen for convenience at the same position $x$ and  
described in the limiting case by a {\it pseudopotential} term  
$W(x)$, depending on pseudospin    
(sublattice)   
indices   
and   
valley (Dirac point)  
indices.     
In this way, the problem of describing line defects in planar systems  
can be mapped to a delta-function pseudopotential $W(x)=W\delta(x)$,  
and this helps us to find    
exact analytic solutions in a simple way. Note, in particular, that  
the pseudopotentials in the form of delta-function barriers with  
Pauli- matrix coefficients, mimicking the   
pseudospin and valley structure of the   
defect line, will be considered as limiting cases of   
induced gauge fields    
arising  
due to perturbations in the hopping parameters  
\cite{Geim0}, \cite{Neto2}, \cite{Vozmediano}.   
 
\section{Pseudopotential for the  effective  2D Dirac equation}     
 
Consider  
a planar system modeling monolayer graphene  
with  electrons in $D=2+1$ space-time.     
Different physical mechanisms give rise to     
(perturbative) interaction terms  
in the effective Dirac  
Hamiltonian that describes electrons in graphene.      
These perturbations may arise  
due to     
several  
types of disorder, like topological lattice  
defects, strains, and curvature.  
Such defects are expected to exist in graphene, as experiments show a  
significant corrugation both in suspended   
samples, in samples deposited on a substrate, and  
also in samples grown on metallic surfaces   
(see, e.g., \cite{Vozmediano}, \cite{Guinea}, and references therein).    
Note, in particular, that changes in the distance between the atoms and in  
the overlap between the different orbitals by strain or bending lead  
to changes in the nearest--neighbor ($NN$) hopping or next--nearest--  
neighbor ($NNN$)   
hopping amplitude and this results in the appearance of vector  
potentials $A_x ( \vecr )$, $A_y ( \vecr )$ (this  
coupling must take the form of a gauge field with the matrix structure  
of the Pauli matrices, $\sigma_1$ and $\sigma_2$)   
and     
a  
scalar potential $ V(\vecr)$ in the Dirac Hamiltonian  
\cite{Geim0},\cite{Neto2},\cite{Vozmediano}.   
Moreover,  
in a region of finite mass the Hamiltonian for Dirac  
electrons should include      
a  
${\vec r}-$dependent mass term $t'=v^2_Fm({\vec r})$ ($m({\vec r})$ is the effective  
mass     
with a $\sigma_3$ matrix)    
due to which the electronic  
spectrum will obtain a finite energy gap. Practically, this type of term  
can be generated by covering   
the surface of graphene with gas molecules \cite{gas1}, or by  
depositing graphene on top of boron nitride \cite{gas2, gas3}.   
 
Let us therefore start with the following  
general expression for the Hamiltonian including the  induced gauge potentials   
and a mass term\footnote{Note that the induced gauge field   
${\vec A}=(A_x,\,\,A_y)$ couples as a complex field  
${\cal A}=A_x+iA_y$ to the pseudospin spinor components, whereas the
scalar potential   
$V$ is real. At the other Fermi point one has to take the
complex-conjugate  field   ${\cal A}^*$ \cite{Geim0},\cite{Neto2}.}  
 
\begin{eqnarray} 
{\cal H} &= &\sum_{\tau=\pm 1} \int d^2x \Psi^{\dag}_\tau ( \vecr) \left\{  
   \sigma_1 \left[ - \vf\ i \partial_x - A_x ( \vecr ) \right] +  
\tau\sigma_2 \left[  
     - \vf i \partial_y - A_ y ( \vecr ) \right]\right\}  
  \Psi_\tau ( \vecr) + \nonumber \\  
&+ &\sum_{\tau=\pm 1}\int d^2x \Psi_\tau^{\dag} ( \vecr )\left[t'( \vecr )\sigma_3 +   V(\vecr)\,\,\, I \right]\Psi_\tau ( \vecr ).  
\label{Ha}  
\end{eqnarray}  
 
Here  the spinors in   
the  
2D plane   $\Psi_\tau  
(\vecr)$ ($\tau=\pm1$, $\vecr=(x,\,\,y)$) have two components   
\begin{equation}  
\Psi_\tau(\vecr)=  
\begin{pmatrix}  
\Psi_{1,\tau} \\  
\Psi_{2,\tau} \\  
\end{pmatrix}  
\label{zhutt}  
\end{equation}  
describing electrons at the  
two $A,\,\,B$ sublattices ($i=1,2$);     
$\sigma_i$ are $2\times2$ Pauli-matrices, $I$ is the unit matrix     
and   
$v_F$ is the Fermi  
velocity   
\footnote{  
Our choice of signs in front of momentum and vector potential components  
of the Hamiltonian essentially corresponds to the conventions of   
\cite{Geim10}, \cite{Neto2}. It may differ from that of other  
papers     
due to different initial definitions adopted.     
However,   
the final results do not depend on it.}.   
 
The physical spin of the electrons that is due to  
spatial rotation properties of the electron wavefunction has been  
neglected in our analysis, and the spinor nature    
of the wavefunction has its origin in the sublattice degrees of  
freedom  called   
pseudospin. The subscript $\tau=\pm 1$ stands for the two Fermi points  
$K,\,\,K'$,  
corresponding to valleys at the corners in the first Brillouin zone   
and     
plays the role of a flavor index.   
Besides     
the above      
effective gauge fields  
an effective electrostatic potential  
barrier may also influence the electron propagation in graphene (see,
e.g., \cite{Geim2}). The term that is   
responsible for this   
(and equally magnetic)   
interaction may be included in the Hamiltonian  
just as an electrostatic scalar potential $e\Phi(\vec r)$  
(and vector potential $e{\vec A_{elm}}$).   
The corresponding property of  
``relativistic'' Dirac electrons in graphene is their ability to  
tunnel through such    
a potential barrier with probability one. This is the so called Klein  
tunneling of chiral particles (see, e.g., \cite{Geim2}\footnote{About
  the Klein paradox of   
relativistic electrons, see the original article \cite{Klein}.}). Its  
presence in graphene is undesirable    
for graphene applications to nanoelectronics. In order to overcome  
this difficulty,   
one may generate   
a gap in the spectrum, which is equivalent to the  
generation of a spatial-dependent mass term. Clearly, the simultaneous existence of a  
scalar potential barrier and a vector gauge field   
${\vec  A}=(A_x,\,\,A_y)$ at some spatial regions may   
influence the electron transmission, say in the $x-$direction.   
In order to study the possible joint role   
and      
competition  
of these  
perturbations, we   
combined them in the model Hamiltonian (\ref{Ha}).   
  
In this way, we assume that the motion of electrons is described by the planar  
Dirac equation $H_\tau\Psi_\tau=i\partial_t\Psi_\tau$ with the Dirac  
Hamiltonian operator   
\begin{eqnarray}  
H_\tau &= &  \sigma_1 \left[ -  \vf  i \partial_x - A_x ( \vecr ) \right] +  
 \nonumber \\ &+ &\tau \sigma_2 \left[  
     -\vf  i \partial_y -  A_ y ( \vecr ) \right]+\vf^2  
 m(\vecr)\sigma_3+ V(\vecr)\,\,\,I,   
\label{Ha1}  
\end{eqnarray}  
where  $\tau=\pm 1$ is the valley index.     
The expression Eq.(\ref{Ha1})   
implies that the low-momentum expansion around the other Fermi point with  
$\tau \to -\tau$ gives rise to a time-reversed Hamiltonian.   
Note that the total effect of both valleys, as described in 4-spinor  
notations \cite{Gus} (and references therein), respects time-reversal invariance. Let us now assume  
that the    
considered  
possible defects are lying in the same spatial region taken, for  
simplicity, to have the form of a line   
lying on the y-axis($x=0$). So our study is   
considered as  investigation of a delta-function limit of  more realistic barrier-type  
configurations and may be  
based on the schematic model Hamiltonian\footnote{Assuming that  
 intervalley interactions are small, nondiagonal mixing terms between  
 spinors belonging to different valleys   
are not considered here.}   
\begin{equation}  
H_\tau=- i\sigma_{1}\partial_{x} - i\tau \sigma_{2}\partial_{y} +W_\tau(x),  
\label{1}  
\end{equation}  
where we     
have   
introduced the  {\it pseudopotential} $W_\tau(x)$  
\begin{equation}  
W_\tau(x)=V(x)I-A_x(x)\sigma_1-\tau A_y(x)\sigma_2+m(x)\sigma_3,  
\end{equation}  
which in  a delta-function limit can be written in the form  
\begin{equation}  
W_\tau(x)=W_\tau\delta(x)=(aI    
-b_{1}\sigma_{1}-b_{2}\tau\sigma_{2}+b_{3}\sigma_{3})\delta(x).  
\label{unit}  
\end{equation}  
In Eq.(\ref{1}), and in what follows, the Fermi velocity, with the  
corresponding choice of the units, is supposed to  
be equal to unity, $v_F=1$. The scalar and vector potentials and the  
mass-type term are chosen as   
$V(x)=a\delta(x),\,\,A_x=b_1\delta(x),\,\,A_y=b_2\delta(x)$,   
$m(x)=b_3\delta(x)$, where $a,\,\,b_i\,\,(i=1,2,3)$ are constants that  
describe the interactions of particles in sublattices from either side of the  
line defect and are related to ``hopping parameters'' (see in what follows).  
   
Let us now apply  
the above schematic model     
to graphene with  
line defects,    
arising  
in the form of     
deformations in the  
structure or  
displacements of     
carbon   
atoms of the    
hexagonal   
crystal lattice    
(some  
of these defects were    
e.g.   
described    
in  
\cite{gun,china}).  
 
The (2+1)-dimensional  Dirac   equation for     
the model under consideration   
\begin{equation}  
\left[ i\partial_t + i\sigma_{1}\partial_{x} -  
 \tau\sigma_{2}p_{y} -  
 W_\tau\delta(x)\right]\Psi_\tau=0  
\label{dirac}  
\end{equation}  
has stationary solutions  
\begin{equation}  
\Psi_\tau(\vecr, t)=  
\begin{pmatrix}  
\Psi_{1,\tau} \\  
\Psi_{2,\tau} \\  
\end{pmatrix}  
e^{i p_{y}y}e^{-iEt},  
\label{zhut}  
\end{equation}  
where  
the  
functions $\Psi_{i,\tau}(x),\,\,(i=1,2)$ should be found by  
a  
limiting procedure\footnote{The problem of the solution of  the low-dimensional Dirac equation with  
a delta-function potential was described in  
\cite{delta1,delta2,delta3} (see also the discussion of the problem in  
\cite{dong,falomir,Loewe}).  
There the authors have shown that the definition of  
$\Psi$ on the boundary of the barrier, which corresponds to an integration of  
the $\delta-$function with the prescription in the limit $\epsilon \to 0$  
as follows: $\int_{-\epsilon}^\epsilon dx \delta(x)f(x)= {1\over  
 2}(f(+\epsilon)+f(-\epsilon)),$   
is unphysical, if one considers  
the $\delta$--potential as a limit of the potential  
barrier. In what follows, it will  
become clear  that     
the   
method  
used in the present paper      
can be considered as appropriate for the description of  
the limiting case of the narrow potential barrier, when the width of  
the short-range   
``$\delta-$function'' potential is considered larger or comparable  
with the width   
of the interval where the function $f$ suffers   
a jump. Using this method  our result will be shown to be in full agreement  
with the result of  
the authors of Ref.{\cite{Geim2}} on the Klein paradox in the   
limit of a very high and narrow barrier.} around the defect line $x=0$.  
The solutions of the free Dirac equation for  
$\tau=+1$ (for $\tau=-1$ the    
corresponding  
solution is also easily  
found) in the  
$x<0$ and $x>0$  
regions can be written respectively as (to simplify notations, we  
omit here the subscript $\tau=+1$)  
\begin{equation}  
\begin{pmatrix}  
\Psi_1\\\Psi_2  
\end{pmatrix}_<=  
\begin{pmatrix}  
1\\  
\e^{i\beta}  
\end{pmatrix}\e^{ip_xx}+B  
\begin{pmatrix}  
1\\  
-\e^{-i\beta}  
\end{pmatrix}\e^{-ip_xx},  
\label{<2}  
\end{equation}  
\begin{equation}  
\begin{pmatrix}  
\Psi_1\\  
\Psi_2  
\end{pmatrix}_>=  
C\begin{pmatrix}  
1\\  
\e^{i\beta}  
\end{pmatrix}  
\e^{ip_xx},  
\label{>2}  
\end{equation}    
where $\beta$ is the incident angle of the electron wave with respect to the x-axis,     
$p_x=p\cos\beta,\,\,\,p_y=p\sin\beta$.   
 
\section{Transmission through the pseudopotential}

Let us next    
study the transmission through the defect line in two particular  
cases of main interest:\\ 1) $b_1\ne  
0,\,\,a=b_2=b_3=0$, and 2) $b_1=0$, $a\ne 0,\,\,b_2\ne 0,\,\,b_3\ne 0$.  
 
\subsection{$b_1\ne  
0,\,\,a=b_2=b_3=0$}  
 
The Dirac  
equations now take the form     
\begin{equation}  
\left\{  
\begin{array}{l}   
E\Psi_{1}+i\Psi^{'}_{2}+i\tau p_{y}\Psi_{2}+b_1\delta(x)\Psi_{2}=0, \\  
E\Psi_{2}+i\Psi^{'}_{1}-i\tau p_{y}\Psi_{1}+b_1\delta(x)\Psi_{1}=0, \\  
\end{array}  
\right.  
\label{s1}  
\end{equation}  
where $\Psi_{i}'=d\Psi_{i}/dx$.  
 
Multiplying the first equation by $\Psi_1$, the second by $\Psi_2$  
and then, in order to exclude the $\delta-$function, subtracting the equations  
we obtain   
\begin{equation}  
E(\Psi_1^2-\Psi_2^2)+i(\Psi_1\Psi_2^{'}-\Psi_2\Psi_1^{'})+2i\tau p_{y}\Psi_1\Psi_2=0.  
\end{equation}  
Dividing this equation by $\Psi_1^2$, integrating over $x$ between  
$-\varepsilon$ and $+\varepsilon$ ($\varepsilon\rightarrow0$) and  
assuming that  
the discontinuities of the functions are finite,  
we  
find the first boundary condition  
\begin{equation}  
\left.\frac{\Psi_2}{\Psi_1}\right|^{+\varepsilon}_{-\varepsilon}=0.  
\label{s}  
\end{equation}  
Now divide the first equation in (\ref{s1}) by  
$\Psi_2$,  
the second by $\Psi_1$,  
and  integrate  both equations over $x$ between  
$-\varepsilon$ and $+\varepsilon$ ($\varepsilon\rightarrow0$). The new  
boundary conditions look like  
\begin{equation}  
\left\{  
\begin{array}{l}  
\left.i\log(\Psi_2)\right|^{+\varepsilon}_{-\varepsilon}=-b_1, \\  
\left.i\log(\Psi_1)\right|^{+\varepsilon}_{-\varepsilon}=-b_1. \\  
\end{array}  
\right.  
\label{s2}  
\end{equation}  
Upon substitution of the solution for the free Dirac equation (\ref{<2}) and  
(\ref{>2})  
in (\ref{s}), (\ref{s2}),  
the transmission probability for both values of the valley indices  
$\tau=\pm 1$ is found to be equal to unity  
\begin{equation}  
T_{\sigma_1}=|C|^2=1.  
\end{equation}  
This result can    
easily  
be   
explained from the point of view of the graphene  
structure. In the two-dimensional graphene model the spinor basis can  
be written in the form  
$  
\Psi_{\tau}=  
\begin{pmatrix}  
\Psi_{1,\tau} \\  
\Psi_{2,\tau} \\  
\end{pmatrix},  
$  
where $\Psi_{1,\tau}, \Psi_{2,\tau}$ are  related to the $A, B$ sublattices of graphene.  
The $\sigma_1$ matrix in front of the $\delta(x)$--function in the Dirac  
equation interchanges the $A$ and $B$ sublattice components in the wave  
function. However in  the tight-binding model of graphene one sums all  
terms over one sublattice, either $A$ or $B$ and the corresponding  
nearest neighbors of the other sublattice, so that the graphene model is invariant under the transformation  
$A\to B,\,\,B\to A$. By this reason,  the incident wave propagates without any  
reflection, since in this  
case the potential  
$-b_1\delta(x)\sigma_1$  
does not form any barrier for it.\\  
 
\subsection{$b_1=0$, $a\ne  
 0,b_2\ne 0,b_3\ne 0$}  
 
Consider the more  
general case with $a,b_2,b_3\ne 0$, and only $b_1=0$. The Dirac  
equation (\ref{dirac}) now takes the form  
\begin{equation}  
\left[E + i\sigma_{1}\partial_{x} - \tau\sigma_{2}p_{y} - \delta(x)(aI-b_{2}\tau\sigma_{2}+b_{3}\sigma_{3})\right]\Psi_{\tau}=0,  
\end{equation}  
which transforms to the set of equations     
(omitting the index $\tau$  
in the wave function)  
\begin{equation}  
\left\{  
\begin{array}{l}  
E\Psi_{1}+i\Psi^{'}_{2}+i\tau p_{y}\Psi_{2}-\delta(x)(a\Psi_{1}+i\tau b_2\Psi_2+b_3\Psi_1)=0,\\  
E\Psi_{2}+i\Psi^{'}_{1}-i\tau p_{y}\Psi_{1}-\delta(x)(a\Psi_{2}-i\tau b_2\Psi_1-b_3\Psi_2)=0.\\  
\end{array}  
\right.  
\label{s3}  
\end{equation}  
 
After    
performing   
some     
further  
transformations   
and    
subsequent  
integration    
in the above  
equations   
to avoid   
problems with the delta-function   
(analogously to what has been done in  
the previous Section),   
we arrive at  the expression  
\begin{equation}   
\left.{1 \over  
   N}\arctan\left[\frac{1}{N}\left((a+b_3)\frac{\Psi_{1}}{\Psi_{2}}+i\tau  
     b_2\right)\right]\right|^{+\varepsilon}_{-\varepsilon}=i,  
\label{sol}  
\end{equation}  
where $N=\sqrt{b_3^2+b_2^2-a^2}$.\\  
  
Substituting $\frac{\Psi_{1}}{\Psi_{2}}$ with the wave functions   
$\Psi_i(x=\pm\varepsilon)$  
from Eqs. (\ref{<2}) and (\ref{>2}),  
we find the transmission probability   
for both values of the  
valley index $\tau=\pm 1$ 
\begin{equation}  
T_{I,\sigma_2,\sigma_3}(\tau)=|C|^2=1-|B|^2=  
\frac{1}{\cosh^2N}\,\,\frac{\cos^2\beta}{\cos^2\beta+  
\frac{(a-b_2\tau\sin\beta)^2}{b_3^2+b_2^2-a^2}\tanh^2N}.  
\label{Trans}  
\end{equation}  
For a beam of electrons propagating towards the line defect, the  
scattered electrons will now be valley-polarized. The valley polarization  
being defined as \cite{Gun2}  
\begin{equation}  
P_{\tau}=\frac{T_{(\tau=+1)}-T_{(\tau=-1)}}{T_{(\tau=+1)}+T_{(\tau=-1)}},  
\end{equation}  
thus  takes the form  
\begin{equation}  
P_{\tau}=\frac{2ab_2\sin\beta\tanh^2N}  
{\cos^2\beta(b_3^2+b_2^2-a^2)+(a^2+b_2^2\sin^2\beta)\tanh^2N}.  
\label{Pol}  
\end{equation}  
As can be seen from the above formula, the valley polarization becomes equal to  
zero for the incident angle $\beta=0$.

\subsection{Comparison with other models}  
   
It is instructive to  
compare the results of the  
previous Subsection, where we have admitted three matrix coefficients  
in front of   
the  
$\delta(x)$-function,    
$I, \sigma_2, \sigma_3$, with some special cases considered in  
the literature.\\   
 
{\bf 1.}  {\it Scalar potential barrier} \cite{Geim2}\\  
 
\noindent    
Clearly,  
the unit matrix $I$ in Eqs. (\ref{Ha}), (\ref{unit}), corresponds to   
diagonal   
pseudospin  
transitions $A\to A,\,\,B\to B$ with respect to the defect  
line, i.e. the graphene to the left of the defect line is  
mirror symmetric to the graphene on the right side     
of  
the  
defect line. This corresponds, e.g.,  to the model of graphene with a  
scalar (electrostatic)   
potential barrier of rectangular shape   
considered in Ref. \cite{Geim2}  
\begin{equation}  
V(x)=  
\left\{  
\begin{array}{l}  
V_0, 0<x<D\\  
0, {\rm otherwise}\\  
\end{array}  
\right..  
\end{equation}  
The authors of \cite{Geim2} obtained the transmission probability for this model  
\begin{equation}  
T_{D}=\frac{\cos^2\beta}{1-\cos^2(q_{x}D)\sin^2\beta},  
\end{equation}  
where $q_{x}=\sqrt{(E-V_0)^2/\hbar^2v^2_{F}-k^2_{y}}$.  
Note that, in the limit $D\to 0,\,\,V_0\to \infty$, $q_xD<\infty$,  
where  $D$ and $V_0$ are the potential barrier width and height,  this  
result goes over  
to  our expression (\ref{Trans}) for the transmission probability for  
a delta-barrier,    
if we put $b_2=b_3=0, a\neq0$, with $q_xD=a$  
\begin{equation}  
T_{a}=\frac{\cos^2\beta}{1-\cos^2a\sin^2\beta}.  
\end{equation}

{\bf 2.} {\it Defect line containing pentagonal and octagonal carbon  
 rings} \cite{Lahiri,China2}\\  
 
\noindent  
 
\begin{figure}[h]  
\center{\includegraphics[width=0.9\linewidth]{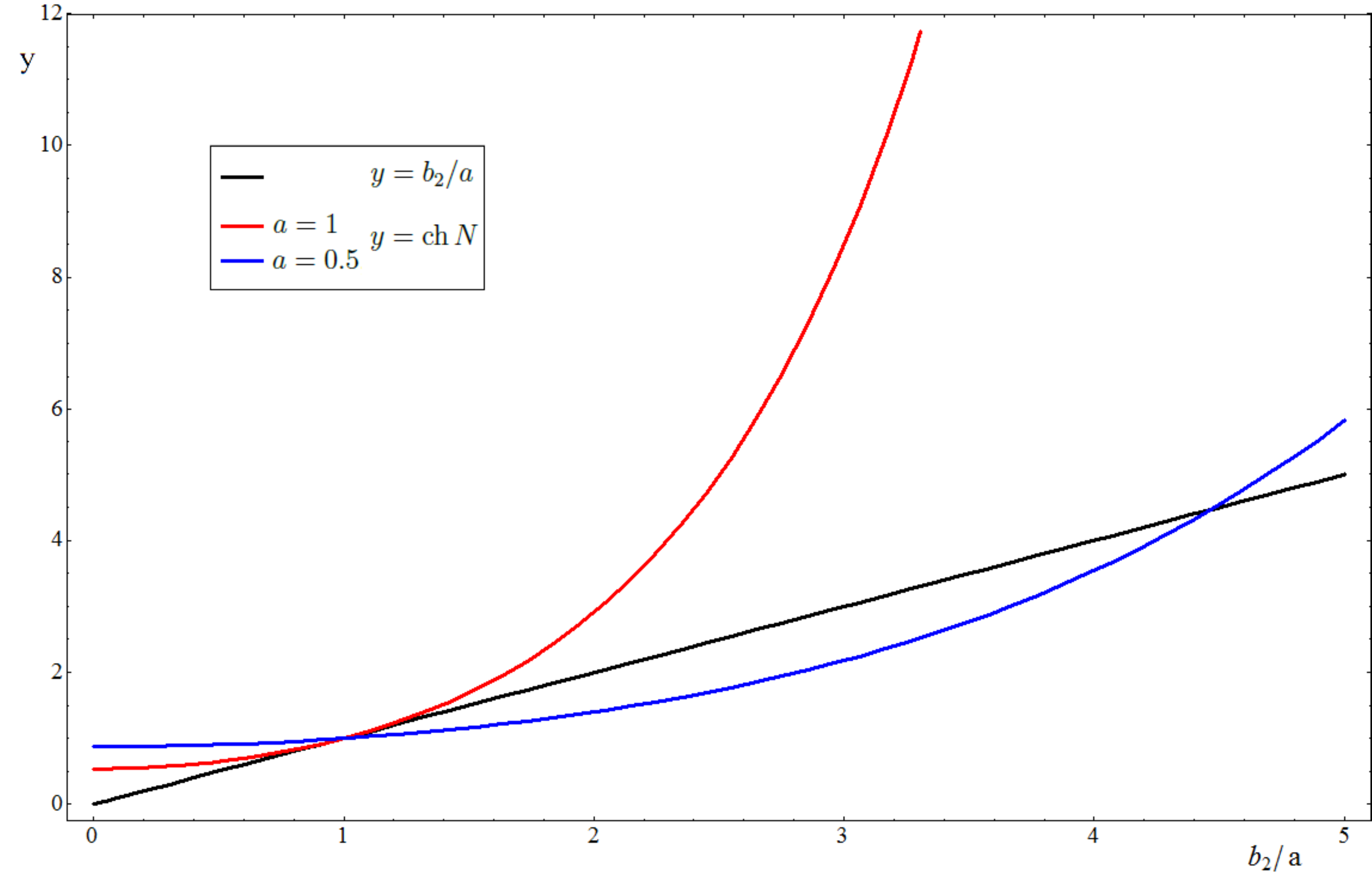} }  
\caption{\small Solution of the equation $\frac{b_2}{a}=\cosh{}N$    
(see (\ref{eq}) in the text)    
for  different   
values of $a$.}  
\label{f1}  
\end{figure}

Let us next consider the model of graphene with a defect line,  
containing pentagonal and octagonal carbon rings, described, e.g. in  
\cite{Lahiri,gun,china,China2}. In \cite{China2} the authors with the use  
of the tight-binding lattice model and the Green function formalism   
obtained the following result for the transmission probability in the  
low energy limit   
\begin{equation}  
T_{(\tau=\pm1)}=\frac{\tau_1^4\cos^2\beta}{(\tau_1^4+\tau_2^2)\mp2\tau_1^2\tau_2\sin\beta},  
\label{Jiang}  
\end{equation}   
where  
$\tau_1,\,\, \tau_2$ are    
NN-``hopping parameters''      
(see Fig.\ref{defect  
 line}).  
Using the notation of the authors $x=\tau_2/\tau_1^2$, one can rewrite  
(\ref{Jiang}) as follows   
\begin{equation}  
T_{(\tau=\pm1)}=\frac{\cos^2\beta}{(1+x^2)\mp 2x\sin\beta}.  
\label{similar}  
\end{equation}  
To compare this expression  
with our results, let us consider Eq.(\ref{Trans})  
in the particular case, when $b_3=0$, i.e. when the    
effective  
mass-type    
term is  
neglected.   
There arises an interesting structural similarity with  
(\ref{similar}),  
if the parameters $a$, $b_2$ of diagonal and non-diagonal pseudospin  
interactions in the pseudopotential (\ref{unit}) are not taken  
independently, but are assumed to satisfy the following relation    
\begin{equation}\frac{b^2_2}{a^2}=\cosh^2(N),  
\label{eq}  
\end{equation}  
where now $N=\sqrt{b^2_2-a^2}$.     
By  
inserting (\ref{eq}) into the  
expression (\ref{Trans}) and putting $b_3 =0$, we obtain in the framework of our    
schematic   
model  
\begin{equation}  
T_{I,\sigma_2}=\frac{\cos^2\beta}{\cosh^2(N)\left[(1+\frac{a^2}{b_2^2})-2\tau\frac{a}{b_2}\sin\beta\right]}=  
\frac{\cos^2\beta}{(1+\frac{b_2^2}{a^2})-2\tau\frac{b_2}{a}\sin\beta}.  
\label{result}  
\end{equation}    
It should be noted that Eq. (\ref{eq})    
has  
besides the trivial solution  
$\frac{b_2}{a}=1$,     
a nontrivial solution for the ratio  
$\frac{b_2}{a}\neq{}1$, if $a<1$. This  can be seen from  
Fig.\ref{f1}. It is clear that for $a\geq 1$ there exists only the trivial  
solution $\frac{b_2}{a}=1$, and the transmission probability     
can reach   
in this case  
its maximum value $T_{I,\sigma_2}=1$ for  
$\beta=\pm {\pi\over 2}$ ($\tau=\pm 1$).  
 
\begin{figure}[h]  
\center{\includegraphics[width=0.35\linewidth]{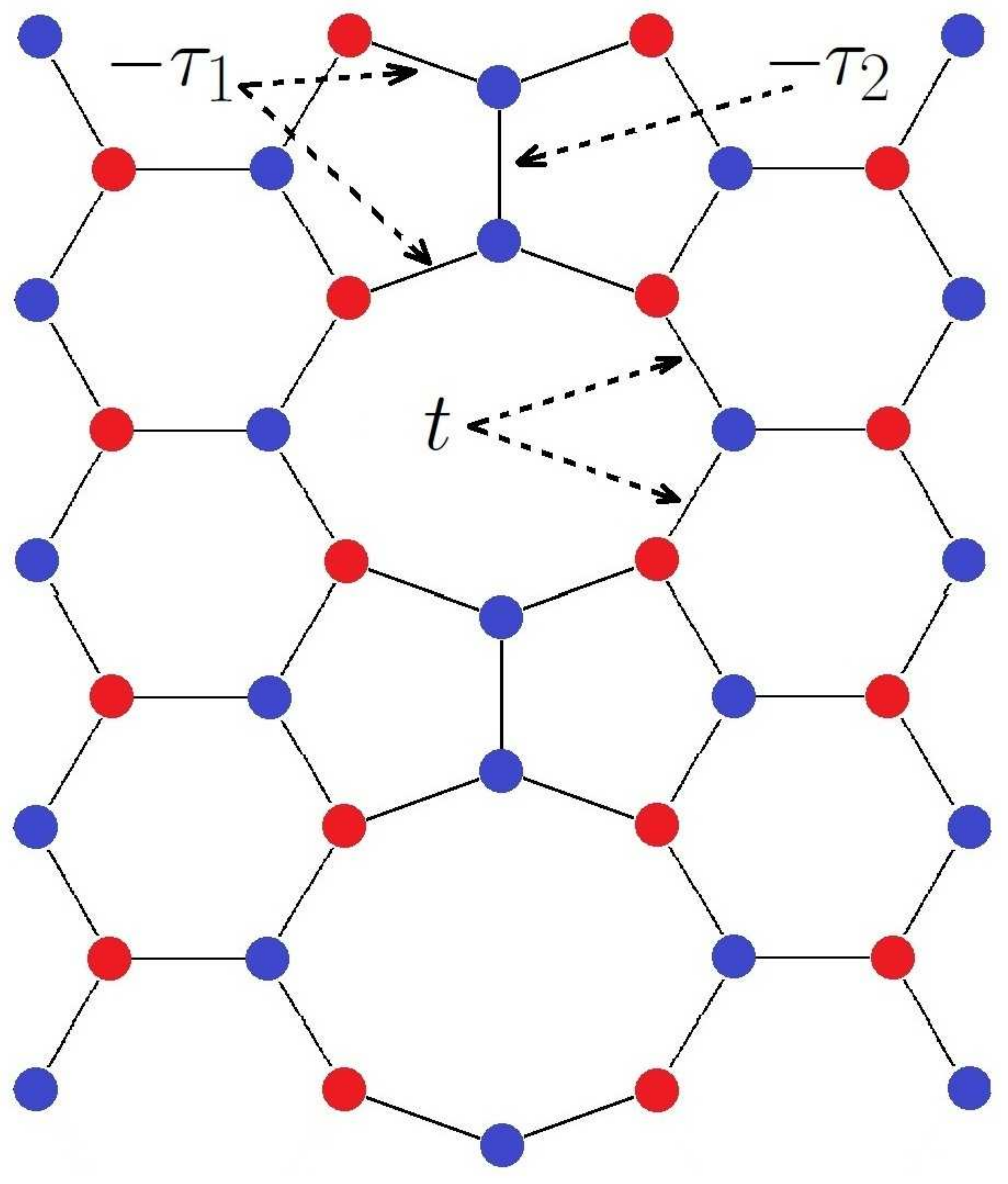} }  
\caption{\small     
Line   
defect, consisting of   
the periodic repetition of one octagonal and two pentagonal    
 carbon rings. Red circles correspond to sublattice A, blue circles  
 to sublattice B; $t$, $-\tau_1$, $-\tau_2$ are NN-hopping energies.  }  
\label{defect line}  
\end{figure}

It is amazing     
to note   
that our result (\ref{result})     
indeed  
looks similar to     
the expression  
(\ref{similar}),     
derived in  paper  
\cite{China2}     
as a low energy limit in a much more involved  
calculation.  
By identifying the expressions $b_2/a=\tau_2/\tau_1^2=x$, it   
thus could be suggested  
that  the  
coefficients $a$, $b_2$  in our pseudopotential  model     
effectively     
correspond to the hopping    
parameter   
quantities  
$\tau_1^2$, and $\tau_2$, respectively. This way, one may   
conclude that $a$ mimics the $NNN$ diagonal     
pseudospin transitions of electrons via two neighboring  
pentagons of the linear defect in Fig. \ref{defect line}, giving  
$\tau_1$-hopping squared,  whereas $b_2$ is responsible for  
$NN$-hopping     
between two mismatched atoms of the B sublattice    
corresponding to $\tau_2$.   
Obviously, such a correspondence between  $a$, $b_2$ and  $\tau_1^2$, and  
$\tau_2$   
supports the original   
interpretation of the role of  
these interactions in the pseudopotential (\ref{unit})    
and looks like a concrete realization of the  
ideas of \cite{Geim0}, \cite{Vozmediano}, where the ``scalar potential'' term  
$a\delta(x)$ mimics the   
$NNN$ hopping,       
while the ``vector-potential'' term  
$b_2\sigma_2\delta(x)$ mimics the $NN$-hopping.    
Note that the above application and interpretation of the pseudopotential method   
required an important additional input: namely, the specific parameter  
relation Eq.(\ref{eq}) which apparently in some effective way reflects the internal  
microscopic structure of the defect line.  
Clearly, such an approach can offer only an  
approximate qualitative description of transmission phenomena.\\   
 
\section{Numerical results}  
 
\begin{figure}[h]  
\center{\includegraphics[width=0.9\linewidth]{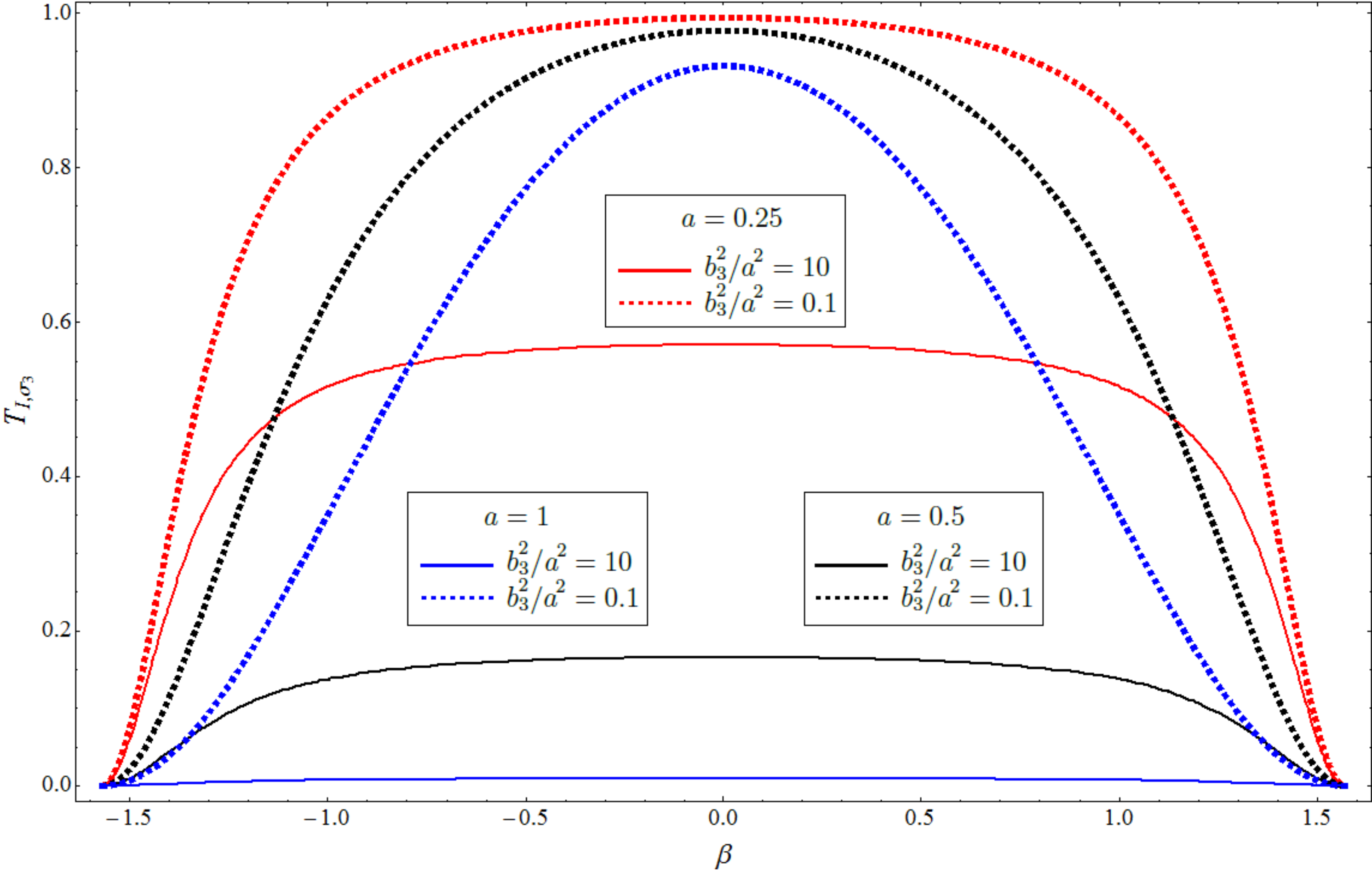} \\ }  
\caption{\small Transmission probability    
$T$    
in dependence of the angle of  
 incidence $\beta$ for $b_2=0$ and for different values of parameter  
 $a$ and ratio $b_3^2/a^2$.}  
\label{b2=0}  
\end{figure}

Let us now return to the expressions for the transmission probability (\ref{Trans}) and  
the valley polarization (\ref{Pol}) of our schematic model for the  
case with $a\ne 0$ and $b_3\ne 0$ simultaneously. The corresponding  
result   
could  be useful for future researches, because the transmission  
probability   
has a nontrivial behavior  (see Fig.~\ref{b2=0}) and the valley  
polarization (\ref{Pol}) equals zero only for zero incident angle  
$\beta=0$.    
 
The  dependence of the transmission on the angle  
of incidence in the case $b_2=0$, $a\neq0$, $b_3\neq0$ is as follows  
\begin{equation}  
T_{I,\sigma_3}=  
\frac{\cos^2\beta}{\cosh^2(a\sqrt{b^2_3/a^2-1})\cos^2\beta +  
 \sinh^2(a\sqrt{b^2_3/a^2-1})\frac{1}{b^2_3/a^2-1}}.  
\label{b}  
\end{equation}  
Its behavior  is shown in Fig.\ref{b2=0}. Obviously, if the  
contribution of the coefficient  
$b_3$ is   
greater   
than that of the coefficient $a$  
the transmission is lower.  
However,  
if the contribution of the coefficient $b_3$ is lower than that of the  
coefficient $a$ for the same values of $a$, the transmission  
probability increases.   
As follows from Eq. (\ref{b}),  
it can reach   
for  values of $a\gg b_3$    
the value   
$T=1$ (for $\beta=0$)   
(see Fig.~\ref{b2=0}).  
 
\begin{figure}[h]  
\center{\includegraphics[width=0.9\linewidth]{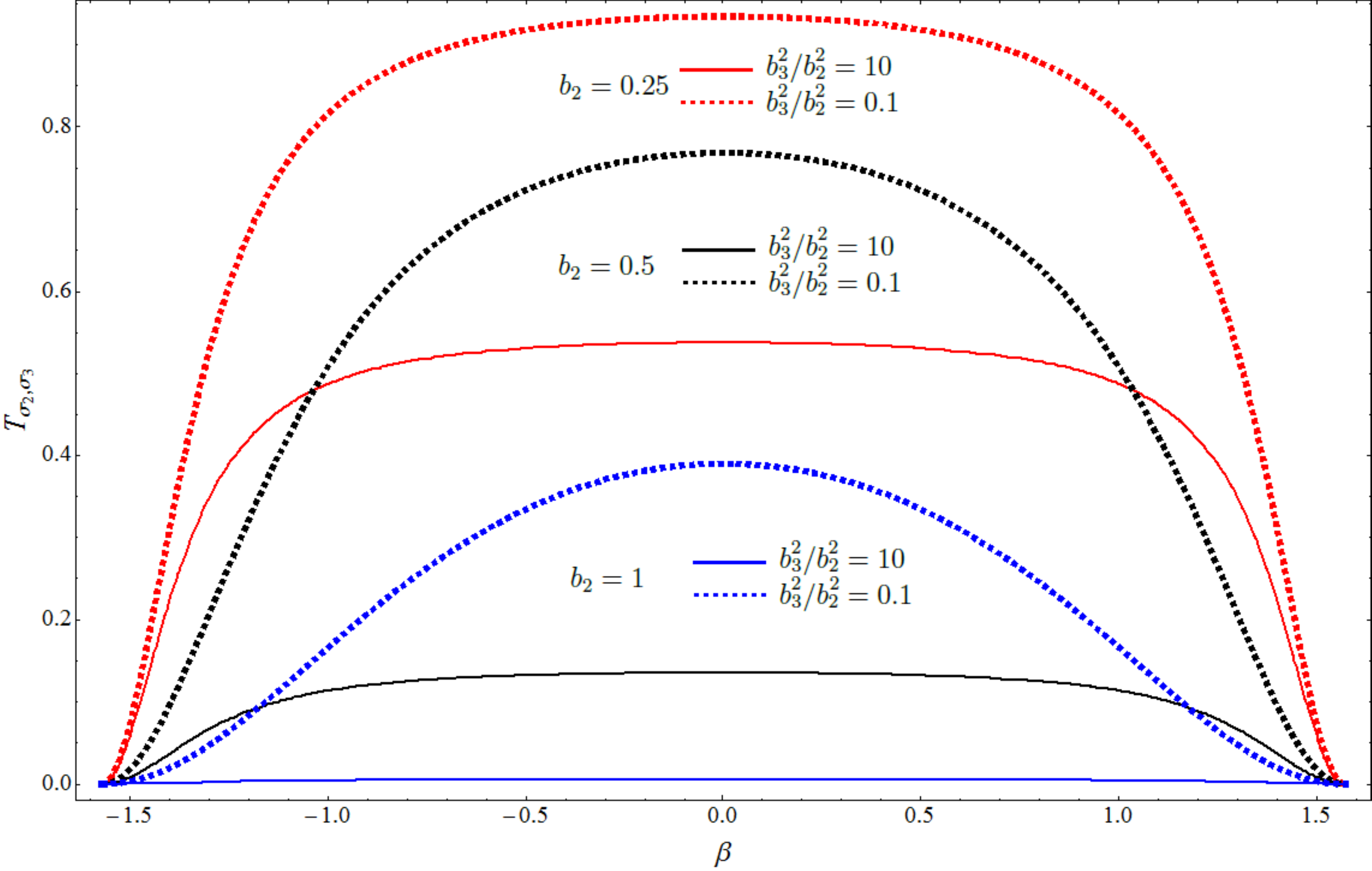} \\ }  
\caption{\small Transmission probability     
$T$  
for $a=0$ and for different  
 values of   
the  
parameter  
 $b_2$ and ratio $b_3^2/b_2^2$.}  
\label{a=0}  
\end{figure}

As is known   
\cite{gas3,Guinea},   
the term in  
the D=(2+1) Hamiltonian (\ref{Ha1}) of the model with the $\sigma_3$ matrix   
corresponds to the effective mass of electrons, and   
as a consequence the electronic  
spectrum will present a finite energy gap.   
The existence of an energy gap  prevents the Klein paradox\footnote{The Klein paradox implies  
that impurities and the other most common sources of disorder will not  
scatter the electrons in graphene.} from  
taking place, a necessary condition for  
building nanoelectronic   
devices made of graphene.  
Our conclusion supports the results of the authors of    
\cite{gas3,Guinea}   
about the role of the mass term as a factor impeding the  
Klein tunnelling of chiral electrons through the barrier.    
The valley polarization for the case $\beta=0$ is still  
equal to zero.  
It should also be  noted that our result (\ref{Trans}), (\ref{b}) for the transmission  
probability in   
the case with only $b_3\neq 0$  corresponds to that of paper  
\cite{gas3} in the limiting case of a narrow  
region with finite mass    
\begin{equation}  
T_{\sigma_3}(\tau)=|C|^2=1-|B|^2=  
\frac{1}{\cosh^2N},\,\,\,N=|b_3|.  
\label{Trans1}  
\end{equation}  
Obviously, in the case $a=0$ the valley polarization (\ref{Pol}) equals zero  
at any angle, and  
the dependence of the transmission probability on the angle of  
incidence is described by the formula   
\begin{equation}  
T_{\sigma_2,\sigma_3}=\  
\frac{\cos^2\beta}{\cosh^2(b_2\sqrt{b^2_3/b_2^2+1})\cos^2\beta +   
\sinh^2(b_2\sqrt{b^2_3/b_2^2+1})\frac{\sin^2\beta}{b^2_3/b_2^2+1}}.  
\end{equation}  
The  
interesting result in this case is that the transmission is  
lower,  
if the  coefficient $b_3$ is greater than the coefficient $b_2$ (see  
Fig.~\ref{a=0}  for various values of $b_2$).  
 
\begin{figure}[h]\centering  
\begin{minipage}[h]{0.9\linewidth}  
\center{\includegraphics[width=0.9\linewidth]{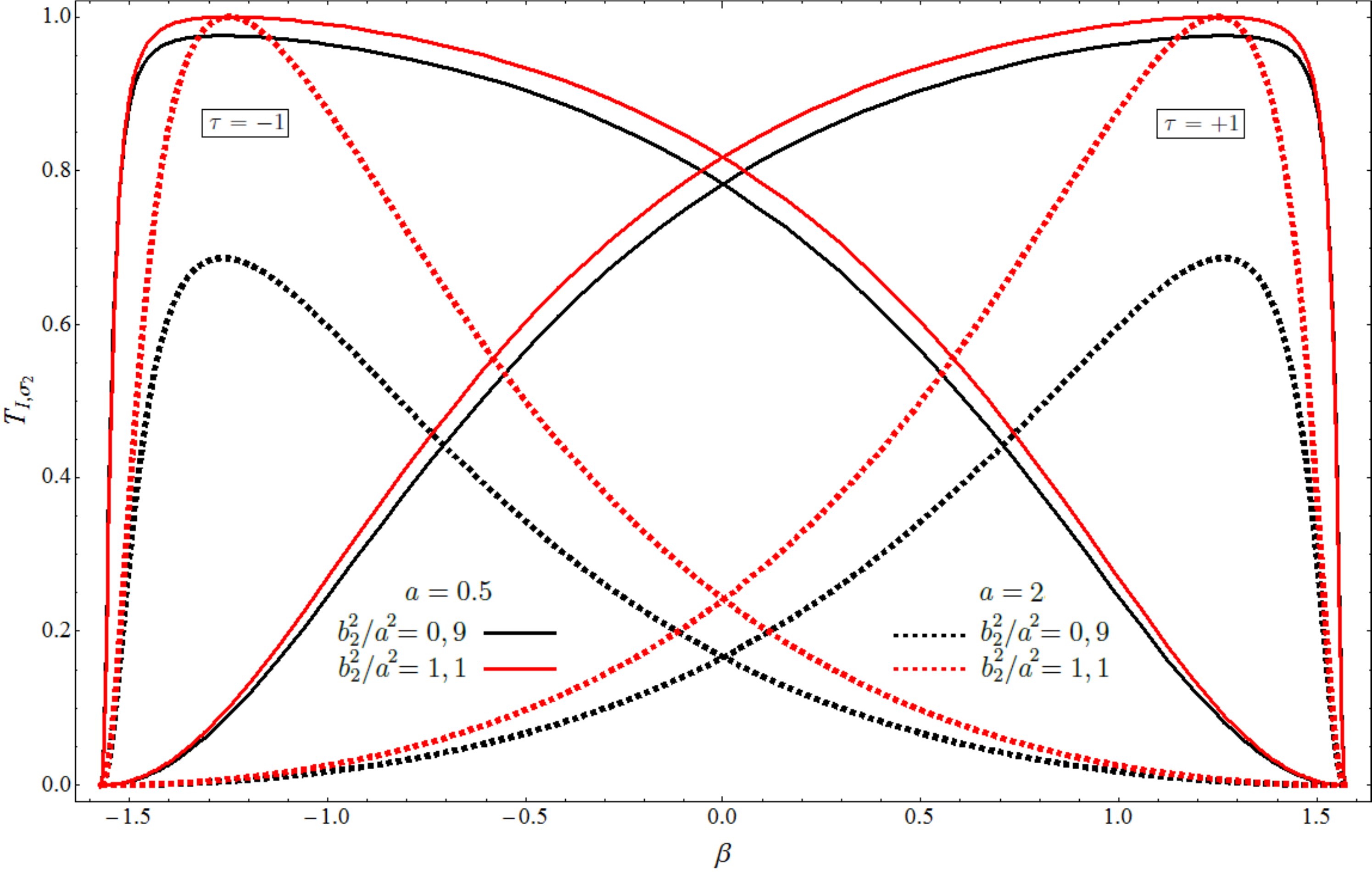} \\ {\small a)}}  
\end{minipage}  
\hfil\hfil  
\begin{minipage}[h]{0.9\linewidth}  
\center{\includegraphics[width=0.9\linewidth]{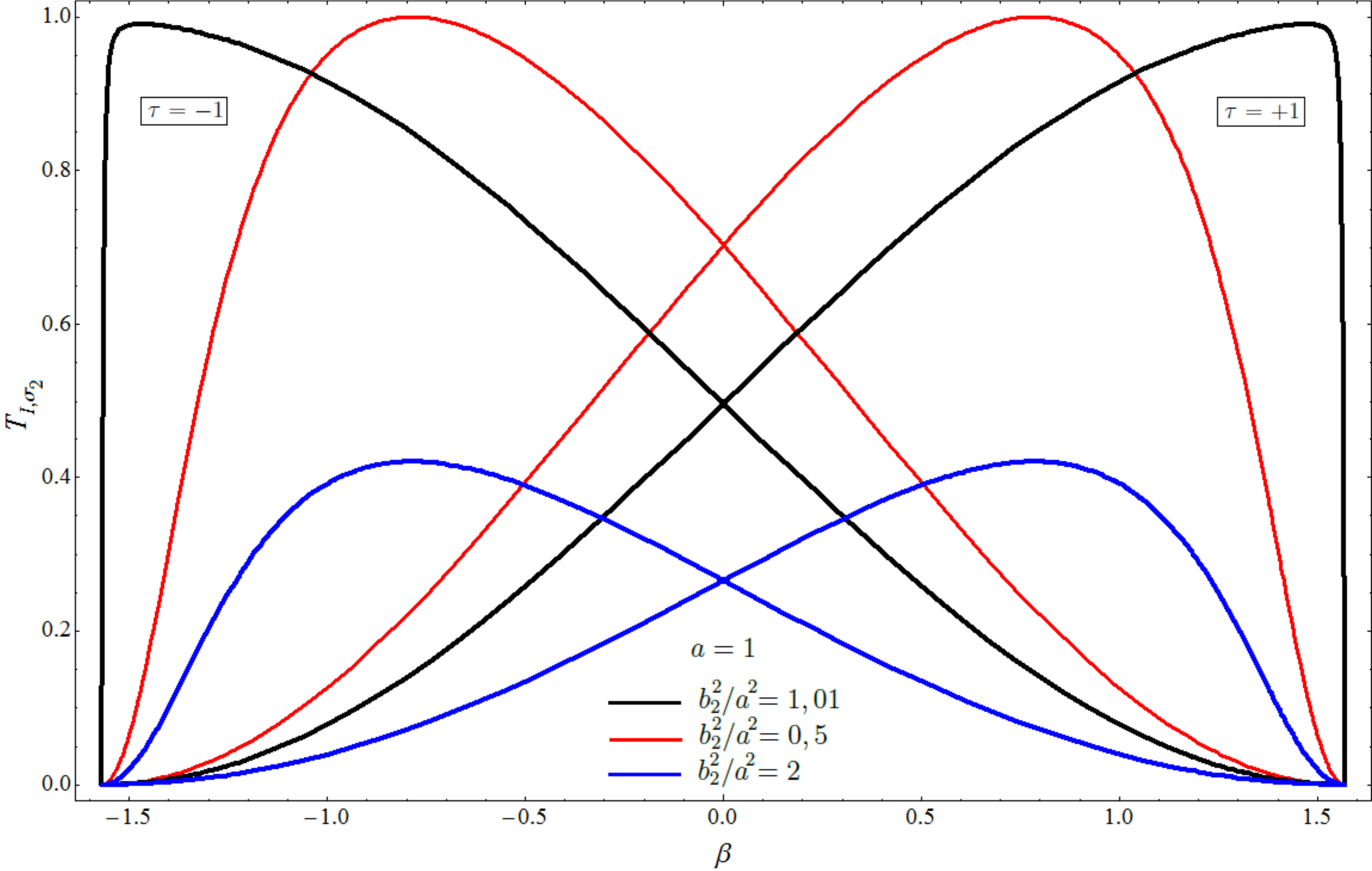} \\ {\small b)}}  
\end{minipage}  
\caption{\small Transmission probability $T$    
for different valley indices $\tau=\pm 1$     
as a function of the  
 incident angle $\beta$ for $b_3=0$ and for different values of   
the  
parameter  
 $a$ and ratio $b_2^2/a^2$.} 
\label{b3=0}  
\end{figure}

\begin{figure}[h]  
\center{\includegraphics[width=1\linewidth]{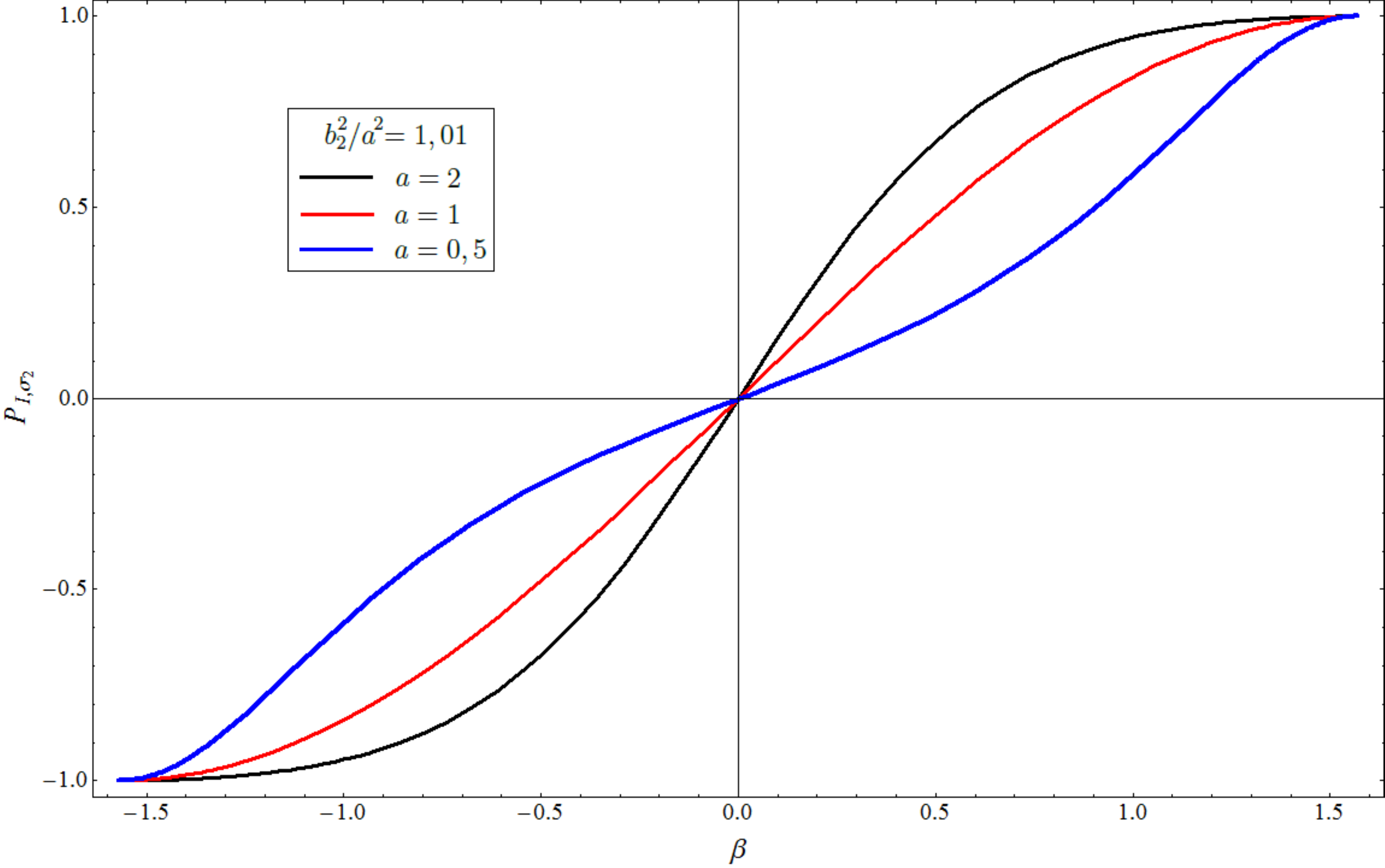} \\ }  
\caption{\small Valley polarization $P$ as a function of the angle of  
 incidence $\beta$  
 for $b_3=0$ and $b_2^2/a^2=1.01$  for different values of   
the  
parameter $a$.}  
\label{P3}  
\end{figure}

\begin{figure}[h]  
\begin{minipage}[h]{0.9\linewidth}  
\center{\includegraphics[width=0.9\linewidth]{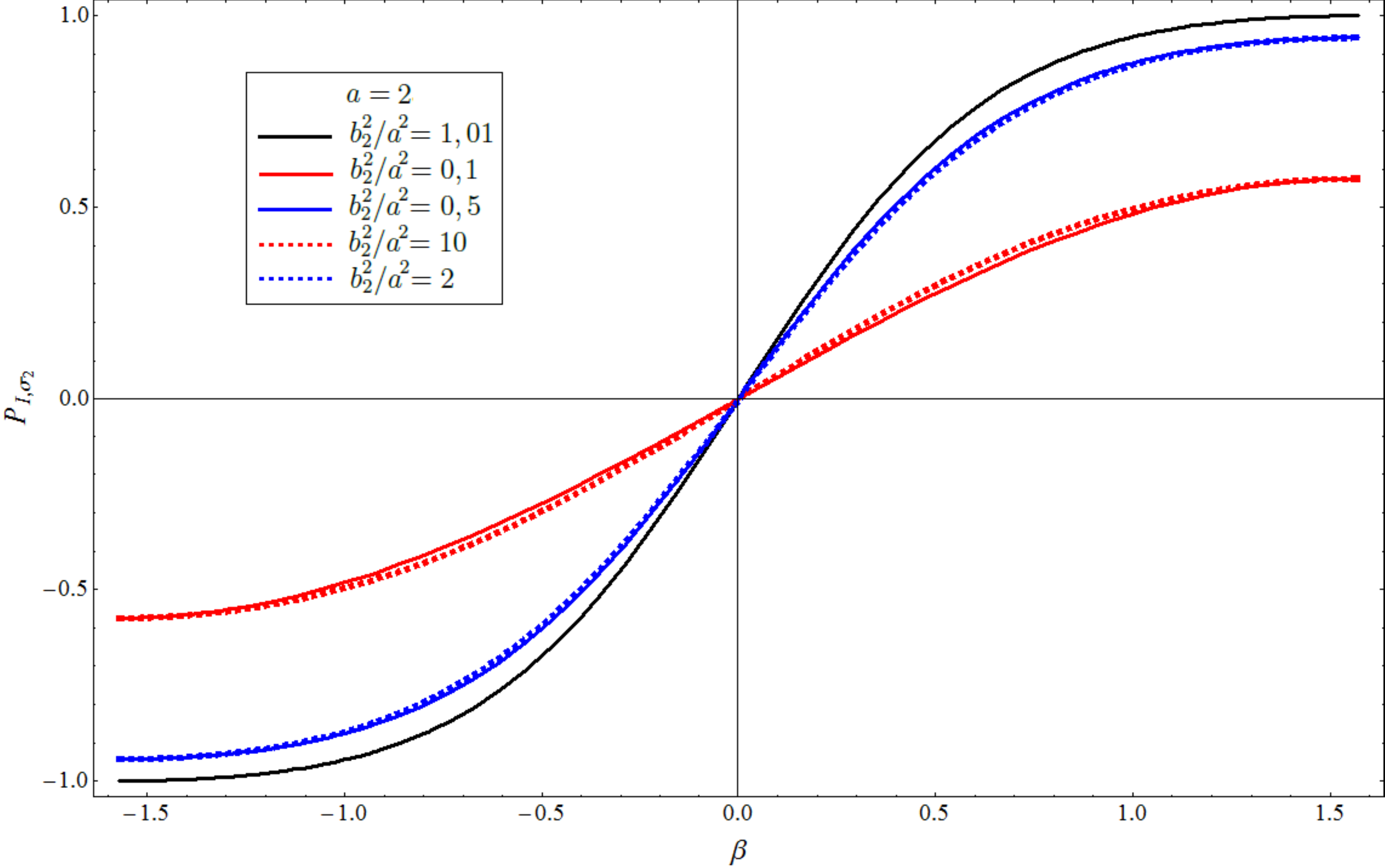} \\ {\small a)}}  
\end{minipage}  
\hfil\hfil  
\begin{minipage}[h]{0.9\linewidth}  
\center{\includegraphics[width=0.9\linewidth]{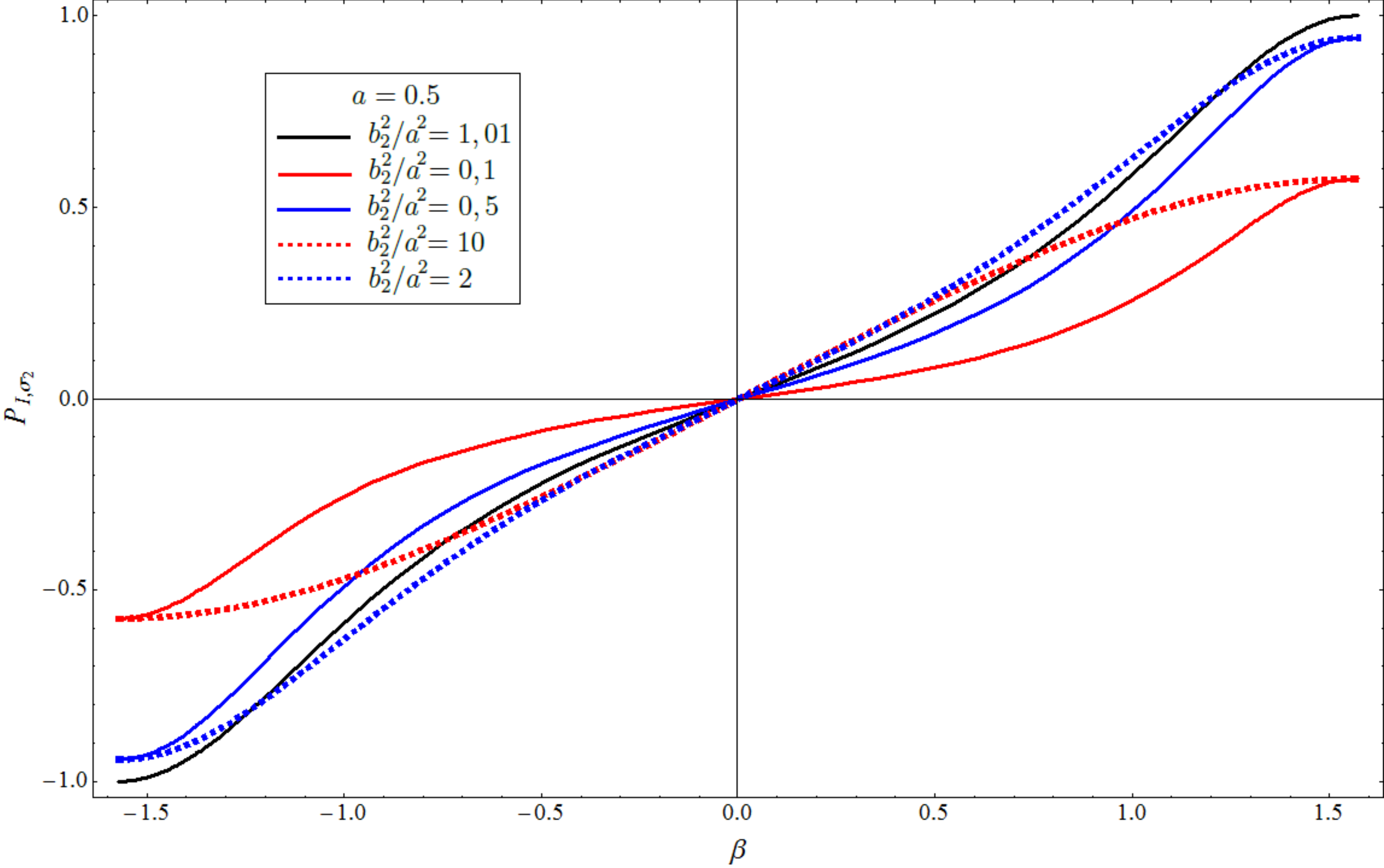} \\ {\small b)}}  
\end{minipage}  
\caption{\small Valley polarization    
$P$   
as a function of the angle of  
 incidence $\beta$ for $b_3=0$ and for different values of   
the  
parameter  
 $a$ and ratio $b_2^2/a^2$.}  
\label{P1}  
\end{figure}

The third case corresponds to  graphene with a  
defect line for $a\ne 0,\,\,b_2\ne 0$ \cite{gun, china, China2}. It  
follows from our general result (\ref{Trans}) that   
\begin{equation}  
T_{I,\sigma_2}=  
\frac{\cos^2\beta}{\cosh^2(a\sqrt{b^2_2/a^2-1})\cos^2\beta + \sinh^2(a\sqrt{b^2_2/a^2-1})\frac{(1-(b_2/a)\tau\sin\beta)^2}{b^2_2/a^2-1}}.  
\end{equation}  
The transmission is still higher for  small values of $a$, and the  
maximum of transmission for the case $a\simeq{}b_2$ is observed for  
the angles $\beta\rightarrow\pi/2$,   
and for the angles $\beta\rightarrow-\pi/2$,  
$(\tau=+1)$ (see Fig.~\ref{b3=0}a)). If the  
contributions of the coefficients $a$ and $b_2$ are not equal  
($a\gg{}b_2$ or $a\ll{}b_2$), the maximum of the transmission probability  
is shifted from the angles $\beta\simeq\pm\pi/2$   
towards   
the center of the  
graph. It should be noted, that the transmission probability for the case  
$a\ll{}b_2$ is small, while for the case  
$a\gg{}b_2$ it tends to $1$ in its maximum for any values of $a$ (see  
Fig.~\ref{b3=0}b)).  
The valley polarization for this case is given in (\ref{Pol}),  
if we set $b_3=0$.     
The result for $a>1$ is similar to that of  paper  
\cite{Gun2} (see Fig.~\ref{P3} black line). The polarization for  
$a\simeq1$ has an almost linear dependence on the angle $\beta$ (see  
Fig.~\ref{P3} red line). However for $a<1$ the dependence of the  
valley polarization on the angle of incidence has  
a nontrivial behavior (see Fig.~\ref{P3} blue line).  
The graphics of the valley polarization for different contributions of the coefficients $a, b_2$ for the same values of $a$ are shown in Fig.~\ref{P1}a),b).  
 
\section{Summary and conclusions}  
In this paper, we have studied a planar electron system for graphene  
containing  a   
defect line with a pseudospin and valley structure by using a  
schematic model with a delta-function pseudopotential.    
The underlying structure of the considered pseudopotential is assumed  
to arise from various perturbations on the line, in particular   
strain, which lead to changes in the $NN$ and $NNN$ hopping amplitudes  
and are represented by vector and scalar    
gauge fields with the matrix structure  
of the sublattice (pseudospin) Pauli matrices and the unit matrix in the  
Dirac Hamiltonian. In addition, a space-dependent mass term, localized  
in  a narrow region of space,  was taken into account and described by  
including a delta-function term   
with a $\sigma_3$ matrix coefficient.    
On this basis, the transmission through a defect line  in the  
graphene structure   
with various pseudospin types of   
defects   
was considered,   
and the transmission  
probability and valley   
polarization were obtained in the framework of the considered schematic model.   
Moreover, we presented  also justifications for dealing with a  
$\delta$--function   
pseudopotential as a   
model of a narrow square barrier by considering  limiting cases  
of special interest.     
Note that in the limit of a narrow square  
barrier   
our calculation proved to be in  agreement  
with the corresponding limit of the result of     
\cite{Geim2} obtained for the    
electrostatic potential  
barrier of finite width  
(Klein paradox).     
Moreover, the considered pseudopotential model allows also an interesting  
effective description of a defect line with linear repetition of two  
pentagonal and one octagonal carbon rings (\cite{Lahiri,China2}).  
In particular, it was shown that our results go over to those obtained  
earlier on the basis of the Green function method (\cite{China2}),  
if the parameters $a$, $b_2$ of diagonal and non-diagonal pseudospin  
interactions in the pseudopotential (\ref{unit}) are not taken  
independently, but are assumed to satisfy the specific relation  
(\ref{eq}).

We hope that the    
considered pseudopotential method and   
results of this paper may help to   
enlarge,  
at least qualitatively,    
our understanding of  
the  transport problems of  
charged   
particles in planar configurations containing line  
defects with various pseudospin structures.

\section*{Acknowledgments}  
Part of this work has been done at the Humboldt University, Berlin. We  
would like to thank  the Institute of Physics   
at HU-Berlin,  and, in particular, its Director, Prof. O. Benson, and also the   
Particle Theory Group  for  
their hospitality. Two of us (E.A.S. and  
V.Ch.Zh.) are grateful   
to DAAD, and one of us (V.Ch.Zh.) also to the Institute of Physics   
at HU-Berlin for  
financial support.

\end{document}